\documentclass[conference]{IEEEtran}
\IEEEoverridecommandlockouts
\usepackage{cite}
\usepackage{amsmath,amssymb,amsfonts}
\usepackage{algorithm}
\usepackage{algorithmicx,algpseudocode}
\usepackage{graphicx}
\usepackage{epstopdf}
\usepackage{subcaption}
\usepackage[hidelinks]{hyperref}
\usepackage[capitalise]{cleveref}
\crefformat{equation}{(#2#1#3)}
\graphicspath{{./figures/}}
\usepackage{textcomp}
\usepackage{xcolor}
\def\BibTeX{{\rm B\kern-.05em{\sc i\kern-.025em b}\kern-.08em
    T\kern-.1667em\lower.7ex\hbox{E}\kern-.125emX}}
    
\usepackage{tikz}
\usepackage{textcomp}
\usepackage{hyperref}
\usepackage{lipsum}

\newcommand\copyrighttext{%
  \footnotesize \textcopyright 2020 IEEE. Personal use of this material is permitted.  Permission from IEEE must be obtained for all other uses, in any current or future media, including reprinting/republishing this material for advertising or promotional purposes, creating new collective works, for resale or redistribution to servers or lists, or reuse of any copyrighted component of this work in other works.}
\newcommand\copyrightnotice{%
\begin{tikzpicture}[remember picture,overlay]
\node[anchor=south,yshift=10pt] at (current page.south) {\fbox{\parbox{\dimexpr\textwidth-\fboxsep-\fboxrule\relax}{\copyrighttext}}};
\end{tikzpicture}%
}
\begin{document}

\title{Line Outage Identification Based on AC Power Flow and Synchronized Measurements
\thanks{This work was supported by the Natural Sciences and Engineering Research Council of Canada (NSERC) under Discovery Grant NSERC RGPIN-2016-06674.}
}

\author{\IEEEauthorblockN{Zhen~Dai,  Joseph~Euzebe~Tate}
\IEEEauthorblockA{\textit{Department
of Electrical and Computer Engineering, University of Toronto,
Toronto, Canada}\\
zhen.dai@mail.utoronto.ca, zeb.tate@utoronto.ca}
}

\maketitle
\copyrightnotice
\begin{abstract}
This paper proposes a method of identifying single line outages in power systems based on phasor measurement unit (PMU) measurements and ac power flow models. 
In addition to the main identification algorithm, a rejection filter is introduced so that the preliminary identified results can be further processed and categorized into three types: correctly identified, misidentified and inconclusive (including correct-filtered and misidentified-filtered).
The methods are systematically tested using test systems of various sizes for various PMU placements, and the results show that the proposed identification algorithm has a high identification accuracy and the proposed rejection filter is able to reduce the misidentified rate without significantly increasing the number of inconclusive cases.
\end{abstract}


\section{Introduction}
Identifying line outages is crucial to understanding current system conditions and preventing cascading outages.
Historically, line outages have been identified based on breaker status indications; however, sometimes the reported line statuses are incorrect. 
Failure to update the system models can cause inaccurate state estimation and threaten system reliability~\cite{NERC_EMSoutage_2017}, especially for systems that are under stress. 
For example, in the 2011 San Diego blackout, operators could not detect overloaded lines because of an incorrect network model~\cite{OutageReport2011,Chen2015Quickest}.

Phasor measurement units (PMUs), because of their unique characteristics (namely, global synchronization and high reporting rate), have been installed on many modern power grids to provide better visibility of grid behavior.
In particular, there have been many reported efforts to use PMUs to improve topology models (in particular, to detect one or more line outages) using various approaches (\cite{Tate_line_2008,Chen2015Quickest,Tate_double_2009,Zhu_sparse_2012,Emami_external_2013}).
Methods in~\cite{Tate_line_2008} and~\cite{Tate_double_2009} are based on hypothesis testing for single-line and double-line outage cases, while reference~\cite{Zhu_sparse_2012} proposes an overcomplete representation and formulates the problem in terms of sparse vector estimation.
Alternative approaches include integer programming (\cite{Emami_external_2013}) and identification based on the small-signal linearized power grid model (\cite{Chen2015Quickest}).
Most of the existing works use PMU voltage angle measurements rather than both magnitude and angle data since they rely on dc power flow models, which only considers angles.
PMU-based disturbance detection methods have been developed and used by utilities and operators.
Line switching events are identified as one of the applications ~\cite{NASPI_using}.
In fact, several applications are being developed and tested (e.g., by ISO New England for external system transmission element tripping and PJM for detection and triangulation of large disturbances).
Meanwhile, Operador Nacional do Sistema Elétrico (ONS) in Brazil deployed a WAMS system which identifies transmission line tripping based on angle disturbance of nearby PMUs.
To the best of our knowledge, details of these applications are not publicly available for comparison.
In this research, we extended the single line outage identification algorithm in~\cite{Tate_line_2008} by utilizing voltage phasor measurements from PMUs.
First, the algorithm computes expected voltage phasors for all possible outage scenarios by solving ac power flows.
Next the pre- and post-outage voltage phasor difference can be calculated (hereinafter called expected values).
By comparing the expected and the observed values, the hypothetical outage scenario that is closest to the observation will be identified. 
Unlike prior approaches, which have relied on the relatively inaccurate dc power flow model, the proposed method uses the full ac power flow model to identify outages.
Ideally, system responses after different outages are distinct enough to be correctly identified.
However, due to measurement uncertainty, if two outages lead to similar (but not identical) responses, they may be confused and thus misidentified.
In some cases, misidentification may lead to wrong operation that aggravates the situation especially when the system is under stressful conditions.
For example, misoperation (tripping heavily loaded but not faulted lines) played a significant role in the 2003 US-Canada blackout and the 2015 Turkish blackout~\cite{abdullah_distance_2018}.
In such cases, a more conservative result (inconclusive) sometimes is better than an incorrect result (i.e., no identification is better than an erroneous identification).
This practical problem has not been addressed in previous research.
In order to acknowledge measurement uncertainty and improve identification accuracy, a rejection filtering technique is introduced.
With the rejection filter in place, instead of definitive identification results, the events are now labeled as conclusive or inconclusive.
In summary, a two-stage framework for single line outage identification is proposed.

Extensive tests have been conducted to show the relationships between identification results and different conditions (including filtering methods, threshold values, and PMU placements) on the IEEE 30-bus system.
Additionally, results using the Ontario network are presented.
Section~\ref{sec: Algorithm} gives the first stage of the algorithm for single line outage identification.
Section~\ref{sec:filtering} discusses the rejection filtering algorithm (the second stage). 
Case studies are presented in Section~\ref{sec:case study}.
Lastly, concluding remarks and future work are presented in Section~\ref{sec: conclusions and future work}.

\section{Stage 1: Main Identification Algorithm}
\label{sec: Algorithm}
In this section, the main identification algorithm will be presented. 
The assumptions we make include: \text{(1)} all measurements are taken from a system in quasi-steady state, and \text{(2)} the ac power flow model is available, since steady-state models are available in modern energy management systems for power flow calculations, such as contingency analysis~\cite{wu_power_2005}.

Inspired by~\cite{Tate_line_2008}, the algorithm is based on hypothesis testing using voltage phasor measurements.
By comparing the simulated voltage changes due to each hypothetical line outage to the observed, the case that is closest to the observations is identified as the outage source.
Outages are assumed to be equally likely in all lines (that would not lead to islands).
\Cref{Eqn: criterion} to \Cref{Eqn: V obs} describe the identification algorithm. 
\begin{align}
\label{Eqn: criterion}
l^{\ast}&=\text{arg} \underset{l \in \mathcal{L}} {\: \min} \:E(l)\\
\label{Eqn: E def}
E(l)&=\|\Delta\boldsymbol{\bar{V}_{exp,l}}-\Delta\boldsymbol{\bar{V}_{obs}}\|\\
\label{Eqn: V exp}
\Delta\boldsymbol{\bar{V}_{exp,l}}&=(\boldsymbol{\bar{V}_{exp,l}}-\boldsymbol{\bar{V}^{pre}})\\
\label{Eqn: V obs}
\Delta\boldsymbol{\bar{V}_{obs}}&=\boldsymbol{\bar{V}_{obs}^{post}}-\boldsymbol{\bar{V}_{obs}^{pre}}
\end{align}
An error measure $E(l)$ given a hypothetical outage in line $l$ (where ${l \in \mathcal{L}}$) is defined as (\ref{Eqn: E def}) to quantify the difference between the expected voltage phasor change and its observed counterpart.
Similar to the list commonly used in online contingency analysis~\cite{savulescu2014real}, the set $\mathcal{L}$ represents all lines to be checked for outage occurrence, the cardinality of which (denoted by ${L}$) is the number of lines to be checked.
The most likely line ${l^{\ast}}$ that leads to the smallest ${{E}(l)}$ (\ref{Eqn: criterion}) is identified as the cause of outage.
The observed voltage change after an event ($\Delta\boldsymbol{\bar{V}_{obs}}$) is defined as the difference between the observed pre-event voltage ($\boldsymbol{\bar{V}_{obs}^{pre}}$) and the post-event voltage ($\boldsymbol{\bar{V}_{obs}^{post}}$).
Analogously, the expected voltage change (${\Delta \boldsymbol{\bar{V}_{exp,l}}}$) is defined based on the pre-event voltage ($\boldsymbol{\bar{V}^{pre}}$, computed from state estimator) and the expected post-event voltage (${\boldsymbol{\bar{V}_{exp,l}}}$, computed for all potential line outages ${l \in \mathcal{L}}$ through power flow calculations).

Note that $\boldsymbol{\bar{V}^{pre}}$ is used instead of the pre-outage voltage measured by PMUs because (\ref{Eqn: V exp}) can then be updated constantly and does not require detection of an outage.
In contrast, (\ref{Eqn: criterion}), (\ref{Eqn: E def}), (\ref{Eqn: V obs}) and (\ref{Eqn: rank}) are only updated when an outage is detected. 
In our study, the ac power flow method, in particular, the Newton method, is used to solve for ${\boldsymbol{\bar{V}_{exp,l}}}$.
This is the most computationally expensive step.
For example, one successful ac power flow solution takes around 0.1 s for the Ontario system (with 3488 buses, introduced in Section \ref{sec: line out Ontario}) using MATPOWER~\cite{Zimmerman_matpower_2011}.
By comparison, the rest of the outage identification only takes $0.0075$ s due to highly efficient sorting algorithms.  
All identification algorithms were implemented in MATLAB on a computer with an Intel Xeon E5-1607 processor and 8 GB RAM.  
The results can likely be improved using better computation resources and simple parallelism \cite{Jun_1995,roberge_parallel_2017}.
However, simulation of system responses due to hypothetical outages is commonly used in contingency analysis, which reduces the additional computation burden associated with the proposed algorithm.
Note that unlike phase angles that are typically used in previous works, all voltage vectors in our algorithm are phasors (with both magnitude and angle) with a length of $P$, where $P$ is the number of PMUs in a system.
Empirically, the algorithm is based on the following ranking of the error measure ${{E}(l)}$.
\setlength{\belowdisplayskip}{3pt}
\setlength{\abovedisplayskip}{3pt}
\setlength{\belowdisplayshortskip}{3pt}
\setlength{\abovedisplayshortskip}{3pt}
\begin{align}
\label{Eqn: rank}
0\leq E(l^{\ast}=l_{r_1})\leq E(l_{r_2}) \leq \ldots \leq E(l_{r_{L}})
\end{align}
The subscript of $l_{r_i}$ means this line occupies the $i^{\text{th}}$ place in the ranking among $L$ potential candidates.
The line outage ($l^{\ast}$) that leads to the smallest error (i.e., $1^{\text{st}}$ in the ranking) with respect to the observation is identified as the cause. 
Given PMU measurements due to the actual outage of line $l_a$, if  ${l^{\ast} = l_a}$, then the outage is successfully identified.

\section{Stage 2: Rejection Filtering Algorithms}
\label{sec:filtering}
When measurements are ideal without uncertainty, almost all outage cases should be identified correctly even with a small number of PMUs installed in the system. 
Two exceptions would be series lines with zero injection at shared buses and identical parallel lines. 
Given measurements with uncertainties, more outage cases (not the aforementioned situations) are likely to be misidentified.
Measurement uncertainty may stem from measurement errors (e.g., instrument transformers, the A/D converters, and the communication cables~\cite{zhu_enhanced_2006}) or random errors (e.g. unknown fluctuations that can not be captured deterministically in principle).
In our research, we use independent Gaussian distributions for voltage phasor magnitude and angle to model measurement uncertainty.

Whether a case will be misidentified is determined by the system as well as the measurement uncertainty model.
A simple example for a 4-bus system (\cite{Zimmerman_matpower_2011,grainger_power_1994}) is used to demonstrate the impact of uncertainty on identification results.
The system consists of 4 buses and 4 branches with a PMU installed on bus 2 and 1 (the reference bus).
Assuming there is no measurement error on the reference bus, the ranking in (\ref{Eqn: rank}) solely depends on measurements at bus 2.
\begin{figure}[htbp]
	\centering
	\includegraphics[width=\linewidth]{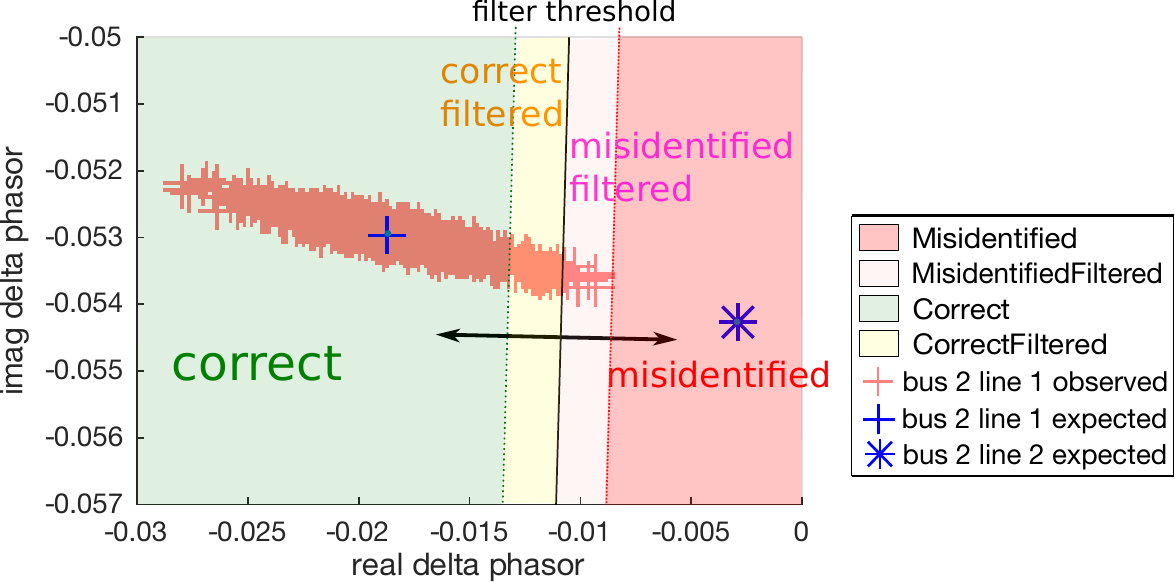}
	\caption[Impact of measurement uncertainty on line outage identification in a 4-bus system]{Demonstration of measurement uncertainty impact on identification, ${\Delta\boldsymbol{\bar{V}_{exp,l}},l\in \{1,2\}}$ and ${\Delta\boldsymbol{\bar{V}_{obs,1}}}$ in 4-bus system, ${\sigma_{v}=0.002/\sqrt{3}}$ pu and ${\sigma_{\theta}=0.01/\sqrt{3}}$ degrees}
	\label{fig_4gs}
\end{figure}
Fig.~\ref{fig_4gs} shows simulated responses after the outage of line 1 (pink crosses) considering uncertainty, along with the ideal response due to the outage of line 1 (blue cross) and line 2 (blue star).
The pink crosses represents 1000 random realizations corresponding to possible measurements after the line 1 outage, generated based on independent distributions.
Specifically, we assume the magnitude and the angle uncertainty follow a zero-mean Gaussian distribution with standard deviation ${\sigma_{v}}$ and ${\sigma_{\theta}}$ respectively.
The main identification algorithm is represented by the solid line that partitions the graph into two regions.
Any point in the left region is closer to ${\Delta\boldsymbol{\bar{V}_{exp,1}}}$ (represented by the blue cross), which means if an observation falls in this region, line 1 will be correctly identified.
However, due to measurement uncertainty, the observations may fall to the right region and be closer to ${\Delta\boldsymbol{\bar{V}_{exp,2}}}$ (represented by the blue star). 
In such cases, the event will be misidentified.

If we set up a rejection filter (represented by the band delineated by two dashed lines on each side of the original identification solid line) so that all cases in the right region are labeled as inconclusive, then they will not be misidentified.
They are in fact misidentified-filtered cases (those would have been misidentified but filtered out to be inconclusive).
However, depending on how we define the filter threshold (band position and width), it is likely that some cases in the left region also fall within the band and become inconclusive (i.e., correct-filtered cases).

There may be different ways to set up the filter threshold. 
In this study, the threshold is determined by $\Delta E$, where ${\Delta E = E(l_{r_2}) - E (l_{r_1})}$. 
Namely the difference between the first two most promising candidates is used to set up the threshold.
\begin{equation}
\Delta E^{(r)}_{l_a} = E^{(r)}_{l_a}(l_{r_2}) - E^{(r)}_{l_a} (l_{r_1}) < \epsilon,
\end{equation}
where ${E^{(r)}_{l_a}(l)}$ denote the error measure in (\ref{Eqn: E def}) but for the actual outage ${l_a}$ specifically.
Due to randomness, multiple runs are conducted based on multiple, independent uncertainty distributions.
The superscript $r$ indicates the error measure is computed based on the $r^{th}$ random realization.

\begin{figure}[!ht]
	\centering
	\includegraphics[width=0.9\linewidth]{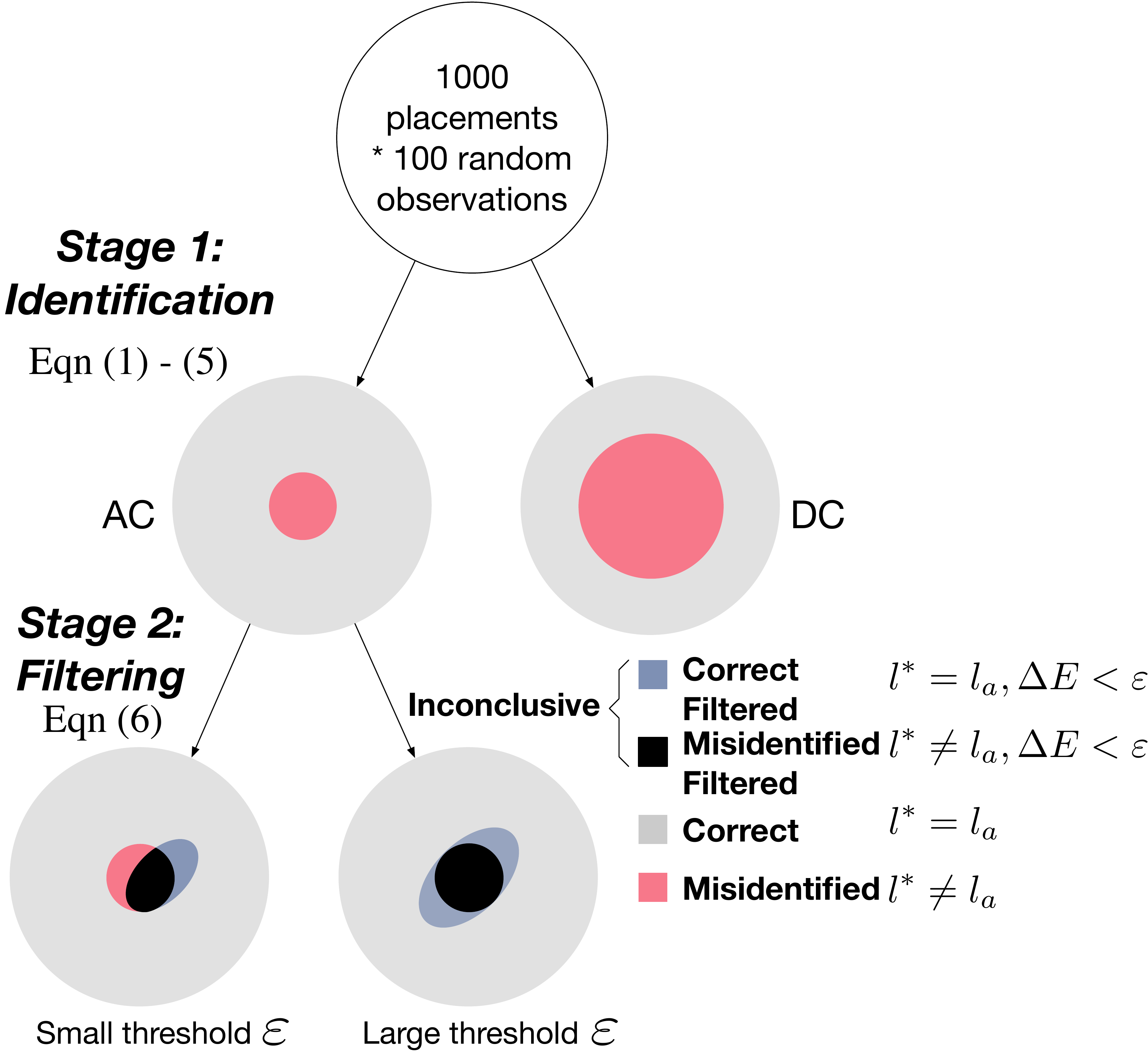}
	\caption{Categories of line identification results (30-bus system with 20 PMUs)}
	\label{fig_results_illustrate_1}
\end{figure}
A visual representation of this process is in \cref{fig_results_illustrate_1}.
The area of each circle represents the number of cases in each category.
The red inner circle represents misidentified cases which correspond to the cases in the right region in Fig.~\ref{fig_4gs} (i.e., misidentified).
If the rejection filter is defined as an ellipse which will not be coincident with the inner circle, then by covering the inner circle (thus eliminating misidentified), some correct cases will become inconclusive (correct filtered) as well.  
The filter threshold defines the sizes of ellipses.
As will be shown in the case studies, as the threshold is increased, the misidentified cases will decrease to zero while the inconclusive cases will go up as a result of increased number of correct-filtered cases.
This corresponds the bottom right circle in Fig.~\ref{fig_results_illustrate_1}. 
In the course of minimizing the misidentified cases, we probably sacrifice some cases which would have been identified correctly. 
The goal is to minimize misidentified cases without introducing a significant number of correct-filtered cases.

\section{Case Studies}
\label{sec:case study}
A comprehensive comparison of different techniques and the choices of filter thresholds are presented in this section. 
The first experiment is the comparison of the ac and the dc approach using the IEEE 30-bus test system. 
Then the results using rejection filtering are presented.
Lastly, the impact of the rejection filter threshold is discussed. 
The identification algorithm is further tested using a 3488-bus model of the Ontario power system.
In all tests, we assume the measurement uncertainty in the magnitude and angle observed follows a zero mean Gaussian distribution~\cite{chakhchoukh_pmu_2014} with standard deviation ${0.002/\sqrt{3}}$ pu and ${0.01/\sqrt{3}}$ degrees respectively.
1000 different placements are generated randomly for each possible ${P\in\{2,\ldots,B\}}$ where ${B}$ is the total number of buses in a system.
If the total number of placements is less then 1000, all placements are considered.
For comparison, 100 randomly generated measurements for each PMU are shared among all placements.

\subsection{IEEE 30-Bus System}
\begin{figure}[hbtp]
	\centering
	\begin{subfigure}[b]{0.48\linewidth}
		\centering
		\includegraphics[trim= 15 0 15 0,width=\linewidth]{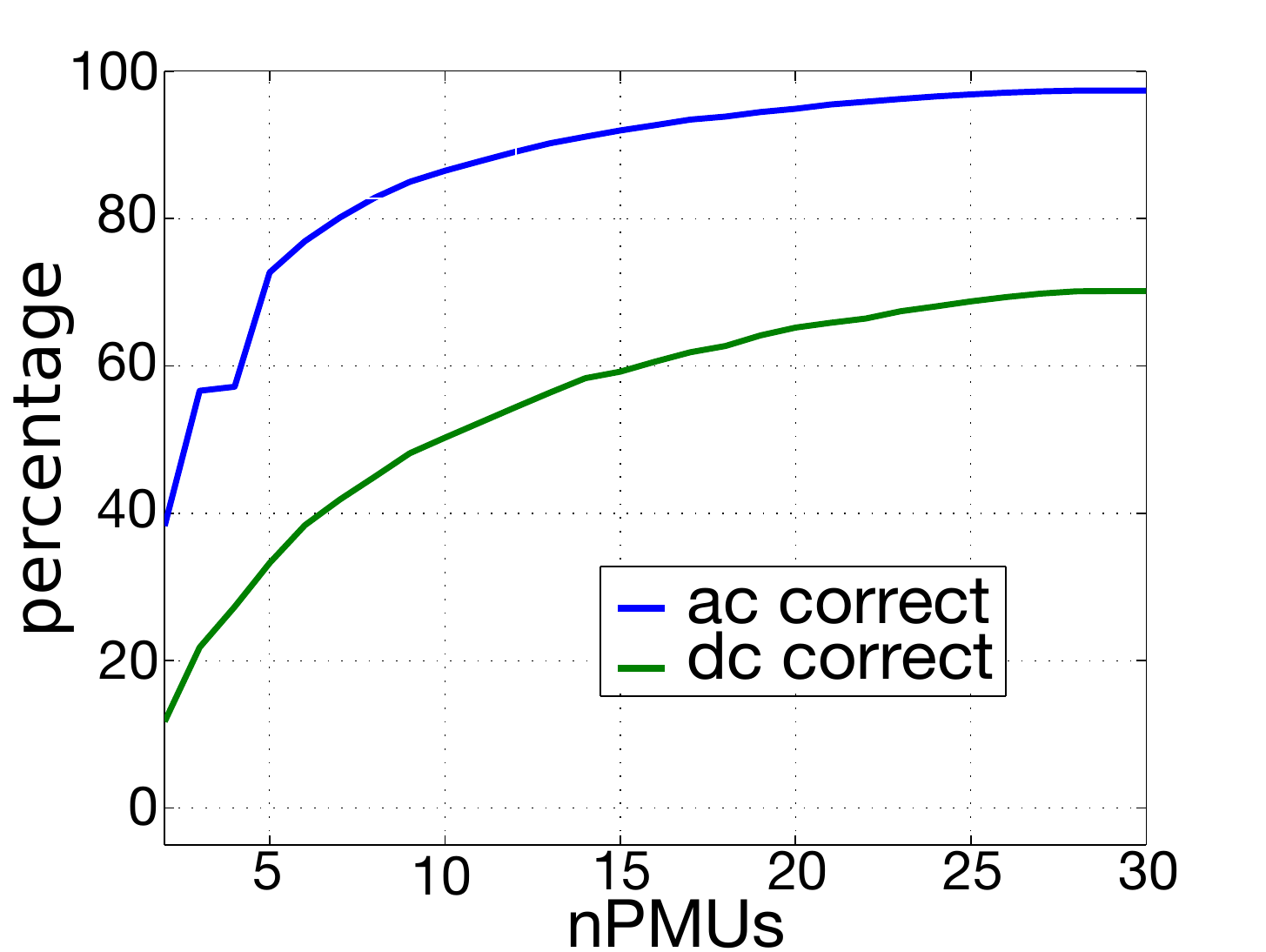}
		\caption{correctly identified, no filter}
		\label{fig_ac_dc_pcorrect}
	\end{subfigure}
	\begin{subfigure}[b]{0.48\linewidth}
		\centering
		\includegraphics[trim= 15 0 15 0,width=\linewidth]{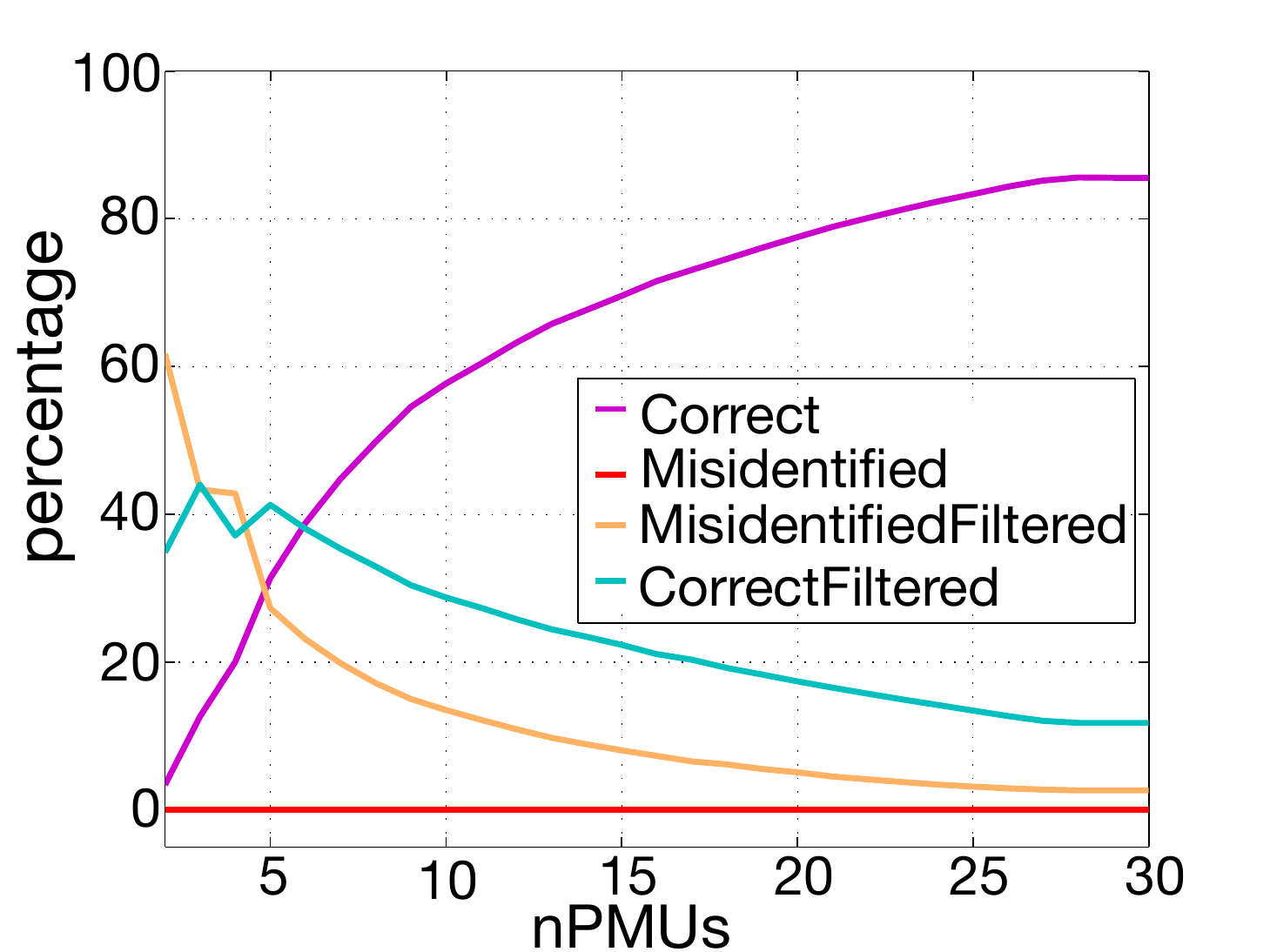}
		\caption{ac, $\Delta E$}
		\label{fig_ac2_set2}
	\end{subfigure}
	\caption[Comparison of dc and ac approaches (${E_{r_2}}$ and $\Delta E$) for line outage identification]{Comparison of dc and ac approaches (with or without $\Delta E$) for line outage identification (The threshold $\epsilon$ is chosen so that the mean misidentification rate is driven under $0.00015\%$)}
	\label{fig: Comparison of DC and 2 AC}
\end{figure}
Fig.~\ref{fig_ac_dc_pcorrect} shows the correctly identified rate of ac and dc approaches without filtering techniques for the IEEE 30-bus system.
As expected, this figure shows that the ac approach is needed to achieve high identification accuracy.
For example, even in the best-case scenario of complete PMU coverage, the dc accuracy is only 70.15\%, whereas the ac accuracy is 97.4\%.  
Fig.~\ref{fig_ac2_set2} shows
performance of the rejection filter based on four types of identification results: correct, misidentified, correct-filtered and misidentified-filtered.
The results in Fig. \ref{fig: Comparison of DC and 2 AC} are generated using thresholds so that the misidentification rate is driven to nearly zero (specifically the mean misidentification rate is driven under $0.00015\%$ over all random realizations, all placements and all outages).
If we compare Fig.~\ref{fig_ac_dc_pcorrect} to \ref{fig_ac2_set2}, the drop of correctly identified rate is because of increased inconclusive cases.
For example, for the ac approach, with 30 PMUs the accuracy was 97.4\%.
It drops to around $83\%$ when the $\Delta E$ filter is used.

\subsection{Results of Different Rejection Filter Thresholds}
This section discusses the impact of the rejection filter $\epsilon$ on the identification results. 
To isolate impact of the threshold, the results should be generated using different PMU placements.
At the beginning of the section, we mentioned 1000 different placements are generated randomly for each possible ${P\in\{2,\ldots,B\}}$ where ${B=30}$.
Additionally, 100 randomly generated measurements for each PMU are shared among all placements.
The number of lines in the 30-bus system to be checked for outages is 38.
If we consider all these possible combinations, $1.102\times10^8$ cases are needed for just one threshold level.
Since enumeration over all possible scenarios takes significant amount of time, we focus on the behavior with a varying range of $\epsilon$ (from 0 to 0.0045) and four specific levels of PMU coverage (10\%, 20\%, 50\% and 100\%).
The measurement uncertainty levels remain the same as in the previous section.

\begin{figure}[!ht]
	\centering
	\begin{subfigure}[b]{0.47\linewidth}
		\centering
		\includegraphics[width=\linewidth]{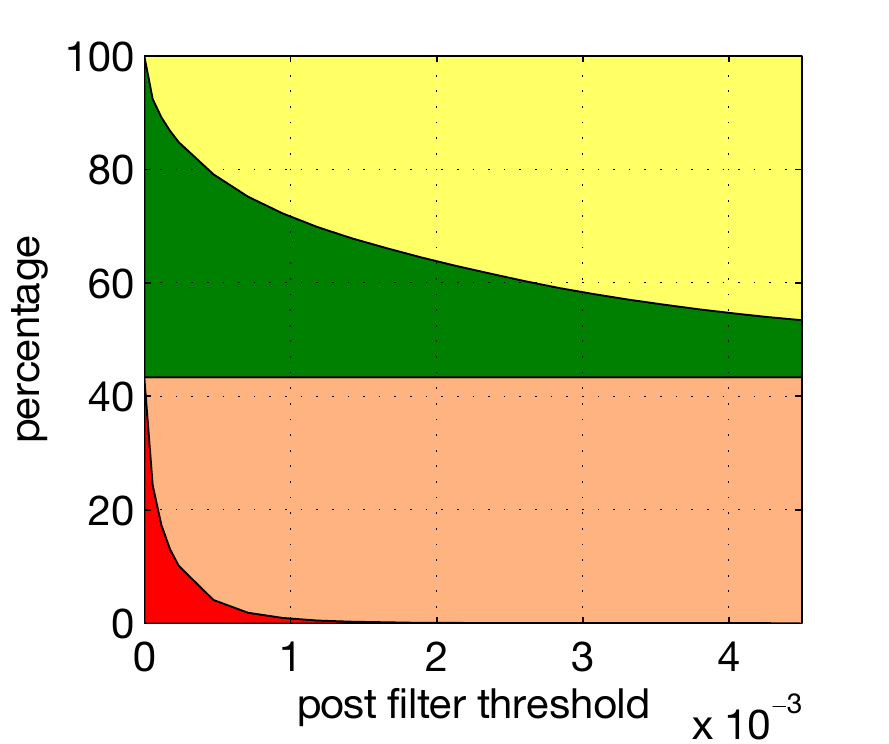}
		\caption{10\% coverage}
		\label{fig_diff_threshold_30bus_10per}
	\end{subfigure}
	\begin{subfigure}[b]{0.47\linewidth}
		\centering
		\includegraphics[width=\linewidth]{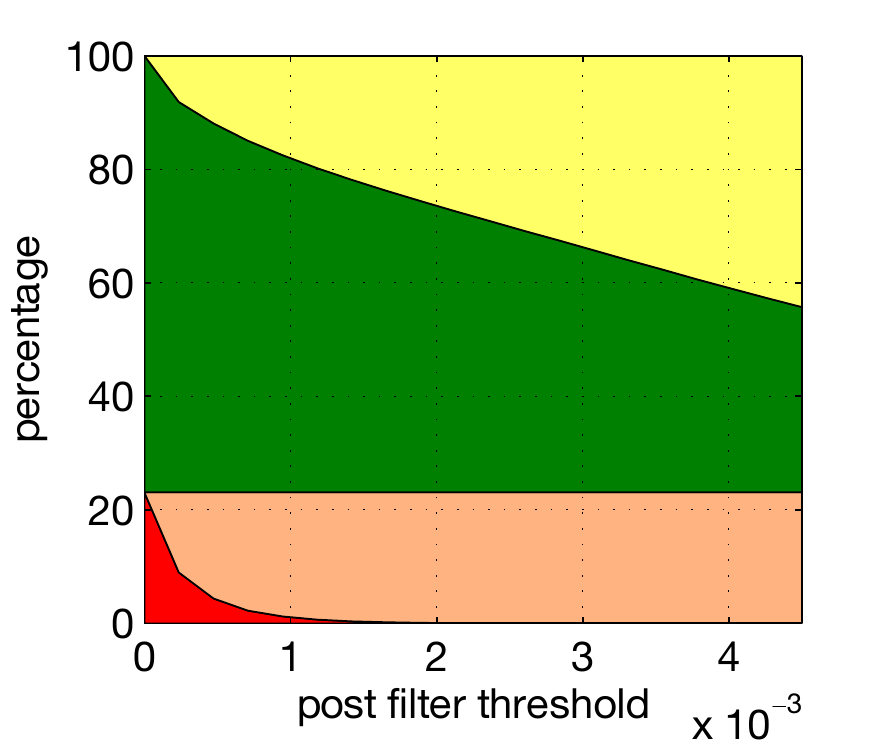}
		\caption{20\% coverage}
		\label{fig_diff_threshold_30bus_20per}
	\end{subfigure}
	\centering
	\begin{subfigure}[b]{0.47\linewidth}
		\centering
		\includegraphics[width=\linewidth]{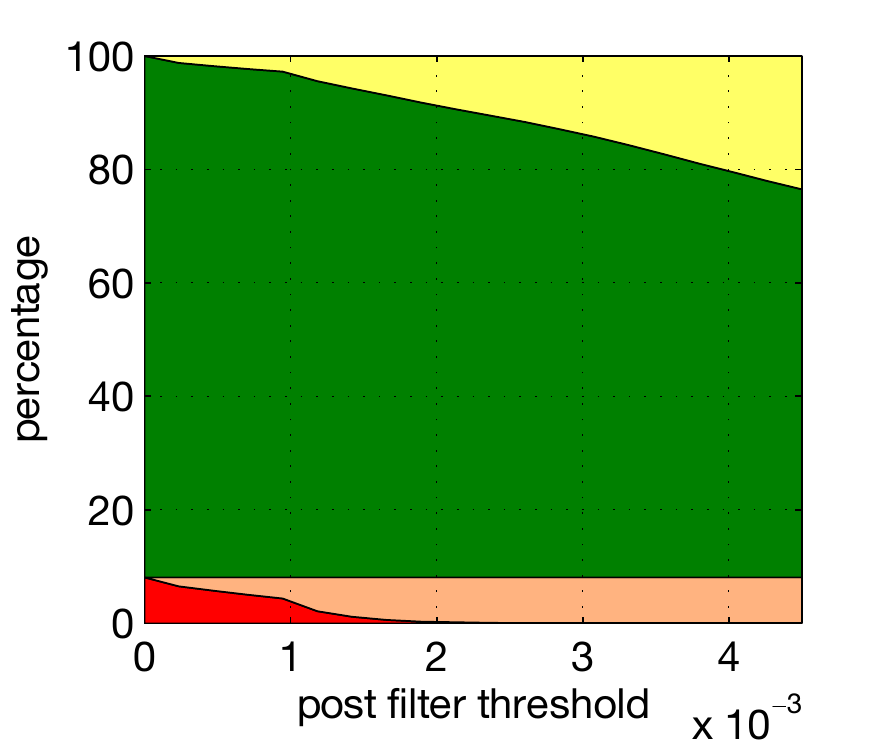}
		\caption{50\% coverage}
		\label{fig_diff_threshold_30bus_50per}
	\end{subfigure}
	\begin{subfigure}[b]{0.47\linewidth}
		\centering
		\includegraphics[trim= 15 0 0 0, width=0.83\linewidth]{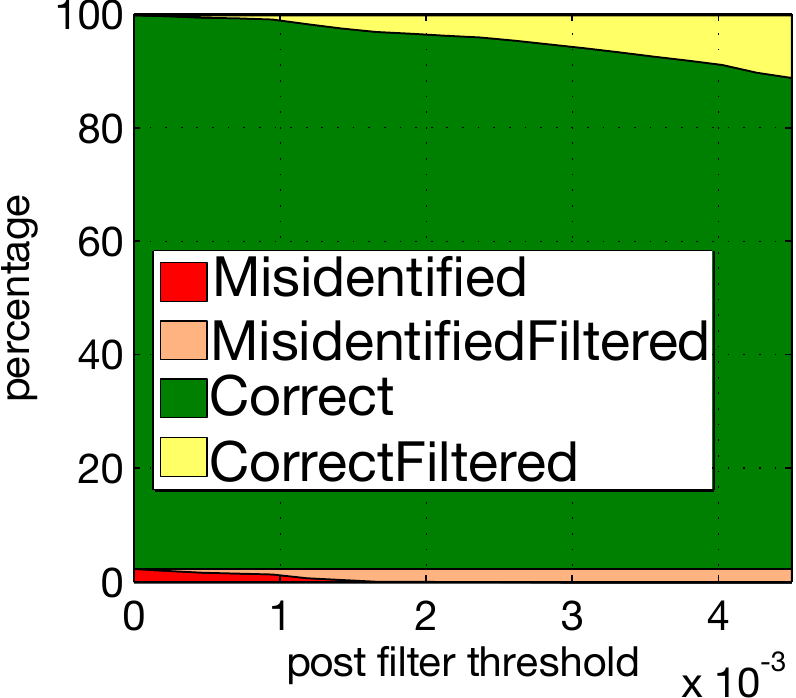}
		\caption{100\% coverage}
		\label{fig_diff_threshold_30bus_100per}
	\end{subfigure}
	\caption[Identification results vs $\epsilon$ (30-bus system, $\Delta E$ filter, four coverage levels)]{Identification results vs $\epsilon$ (30-bus system, $\Delta E$ filter with different coverage (legends are the same for 4 figures))}
	\label{fig_ac_thresholds_complete}
\end{figure}
The trade-off between correct and misidentified cases as a result of different $\epsilon$ is illustrated by Fig.~\ref{fig_ac_thresholds_complete}.
The horizontal lines in each figure represents the correctly identified/misidentified level without filtering.
For example, without any rejection filter, with $10\%$ coverage, the misidentified rate is about $45\%$ while the correctly identified rate is $55\%$.
This corresponds to a red inner circle with an area of $45\%$ of the entire circle in Fig.~\ref{fig_results_illustrate_1}.
By increasing the threshold, more misidentified cases (in red) will be filtered out and become misidentified-filtered (in pink), whilst correctly identified cases (in green) become correct-filtered (in yellow).

Fig.~\ref{fig_ac_thresholds_complete} indicates the misidentified percentage (in red) is very sensitive to the threshold when $\epsilon < 1\times10^{-3} $ (particularly when the coverage is low).
In such cases, the misidentified rate drops rapidly as $\epsilon$ increases from zero.
It also decays faster than the correctly identified rate (by comparing red to green).   
For example, with 10\% coverage (Fig.~\ref{fig_diff_threshold_30bus_10per}), by increasing the threshold from 0 to $1\times10^{-3}$, the misidentified rate drops more than 40\% while the correctly identified rate only drops about 30\%.
However, when the misidentified rate is low (around zero), further elimination of misidentified cases leads to significant loss of correctly identified cases. 
For example, by increasing the threshold from $1\times10^{-3}$ to $2\times10^{-3}$ in Fig.~\ref{fig_diff_threshold_30bus_10per}, the green area decreases significantly (about $10\%$) compared to the trivial gain in eliminating red cases. 
The results also indicate that, ultimately, higher coverage is needed to achieve better identification accuracy.
As the number of PMUs goes up, the ratio between correctly identified and correct-filtered cases also increases (i.e., green versus yellow).
This means fewer correctly identified cases will be sacrificed when the number of PMUs is high. 

Depending on the acceptable level of the misidentification rate or the correct-filtered rate, operators can decide on an empirical value for the threshold.
If misidentification is considered to be worse than a lower correct rate (e.g. compared to false alarms, the operator would rather accept more inconclusive cases), then the threshold can be set accordingly to eliminate misidentified cases.
Otherwise, a very small non-zero level (e.g. $0.0015\%$) can be chosen instead, with minimal impact on the correctly identified cases.

\subsection{Results of the Ontario Power System}
\label{sec: line out Ontario}
To evaluate performance on a more realistic system, results were also obtained using a model of the Ontario power system, consisting of 3488 buses, 864 generators, 1290 loads, 2242 branches and 1697 transformers.
Among the original branches, 1762 single-line outage cases are to be checked for occurrence without introducing islands.
PMUs are assumed to be placed on the actual PMU locations and, alternatively, a set of high voltage buses in Ontario.

First, tests were conducted using 26 bus voltages based on the current PMU locations in the Ontario network \cite{curtis_north_2011}.
To evaluate the performance of the line outage detection algorithms for realistic, future PMU deployments, we also considered cases where all buses above a certain voltage level are monitored by PMUs.
In particular, we consider two cases: monitoring all buses with nominal voltages greater than or equal to 230 kV (an additional 53 buses) or 220 kV (an additional 842 buses).
\begin{figure}[hbtp]
	\centering
	\begin{subfigure}[b]{0.5\linewidth}
		\centering
		\includegraphics[trim=0 0 20 20, height = 0.15\textheight,width=\linewidth]{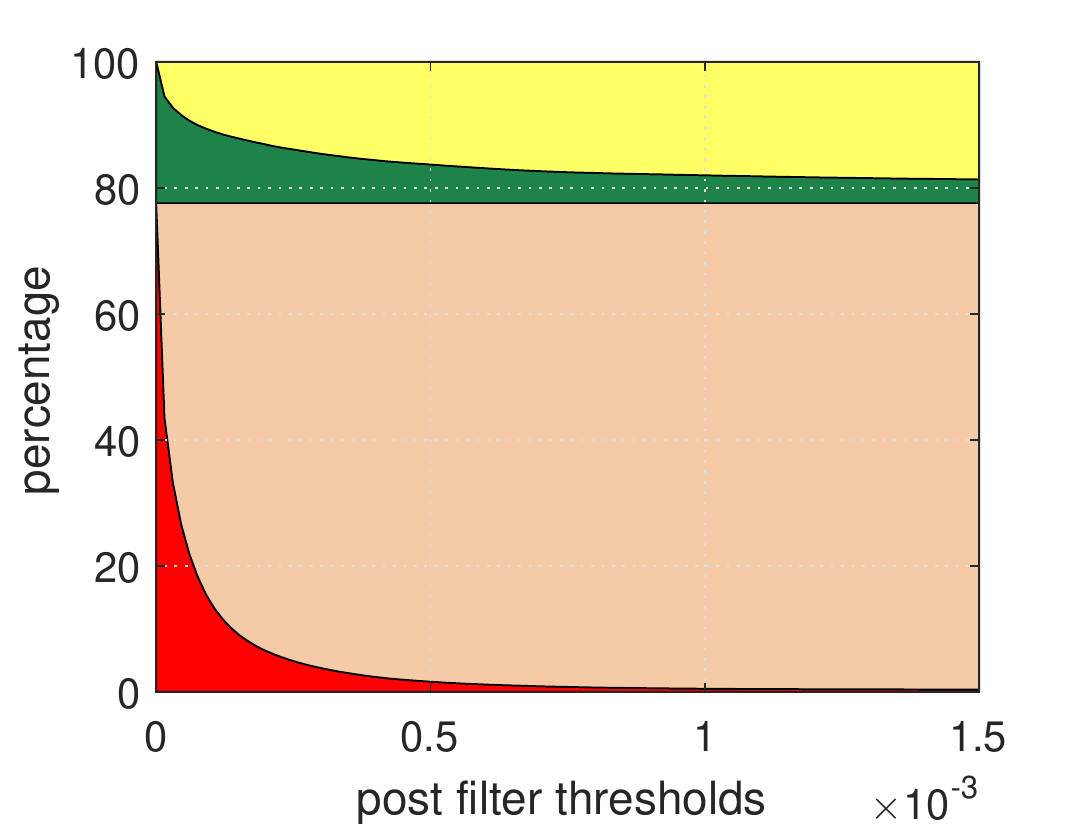}
		\caption{26 PMUs}
		\label{fig_ieso_26PMU}
	\end{subfigure}
	\begin{subfigure}[b]{0.5\linewidth}
		\centering
		\includegraphics[trim=0 0 20 0, height = 0.15\textheight,width=\linewidth]{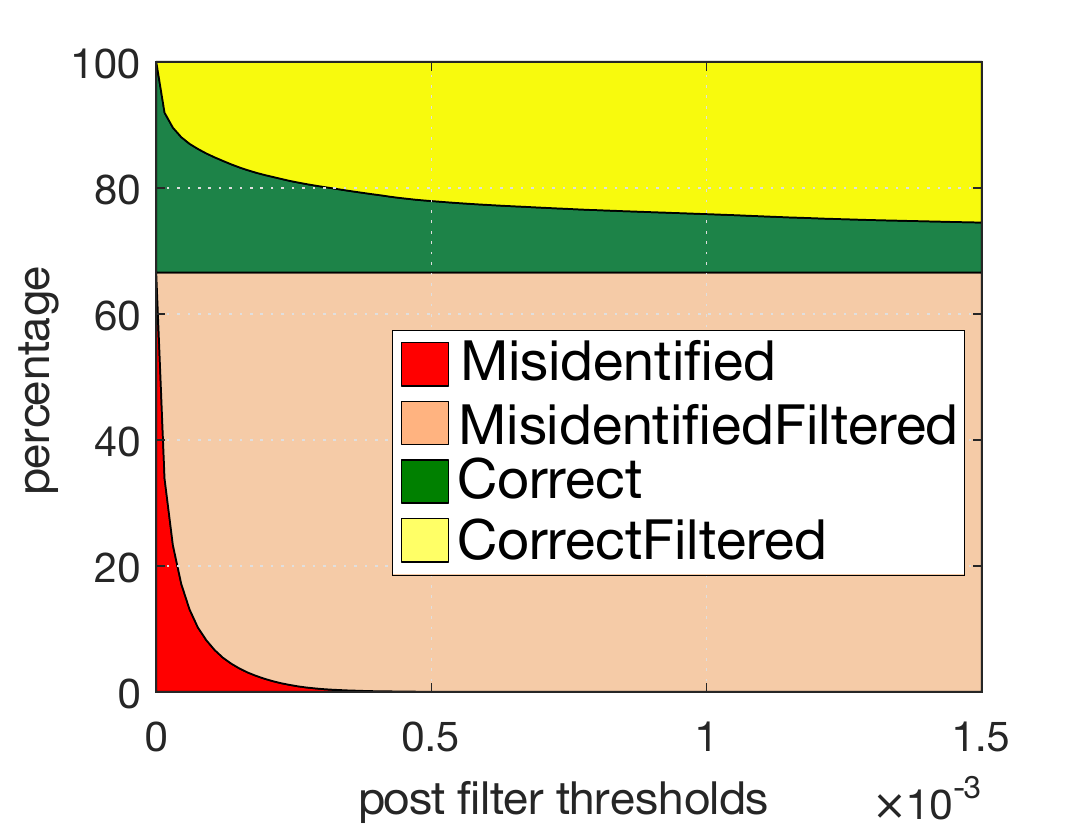}
		\caption{79 PMUs (230 kV+)}
		\label{fig_ieso_79PMU}
	\end{subfigure}
	\begin{subfigure}[b]{0.48\linewidth}
		\includegraphics[trim=0 0 20 0, height = 0.15\textheight,width=\linewidth]{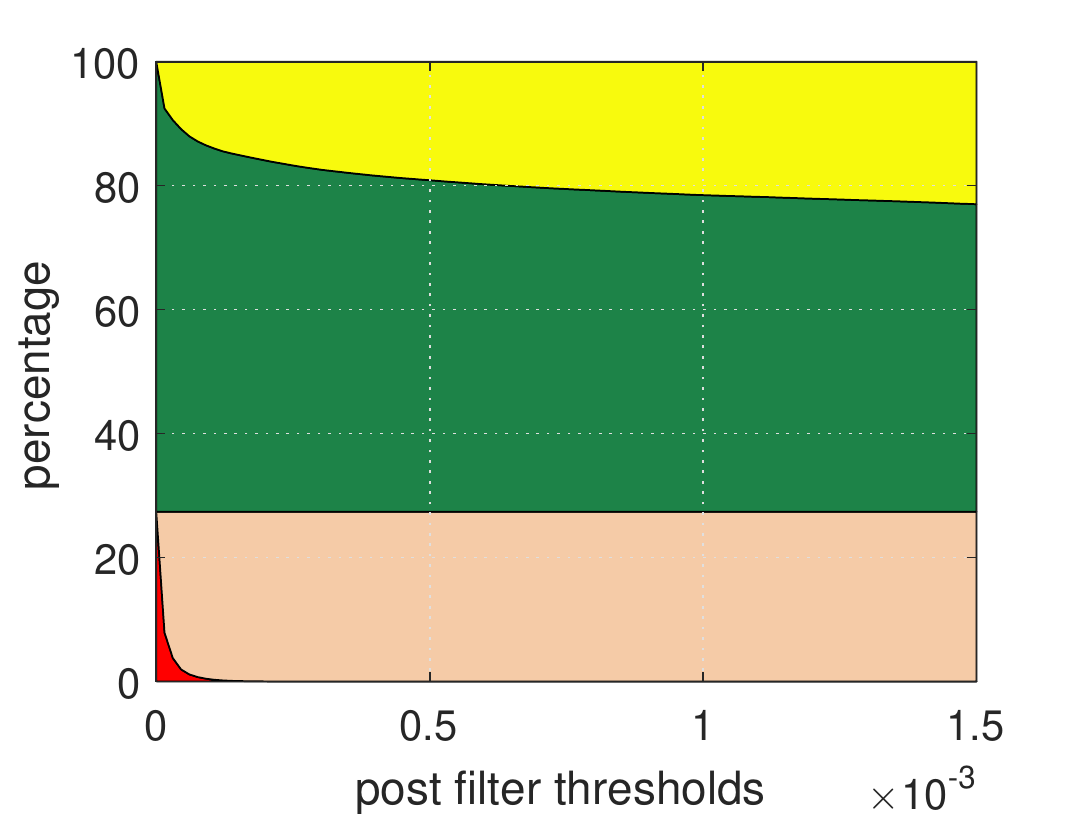}
		\caption{868 PMUs (220 kV+)}
		\label{fig_ieso_868PMU}
	\end{subfigure}
	\caption{Identification results vs $\epsilon$ (the Ontario system, $\Delta E$ filter, three coverage levels)}
	\label{fig_ieso_complete}
\end{figure}
Fig.~\ref{fig_ieso_complete} shows the identification results using different thresholds.
Consistent with the previous results, the observations about impact of filter threshold and PMU coverage are still valid for the Ontario system.
When the coverage goes up, the threshold required to eliminate misidentified cases also decreases, which results in not only a higher correctly identified rate but also higher ratio of correct versus correct-filtered.
The horizontal line in each figure represents the separation of correctly identified (above the line) and misidentified (below the line) without rejection filter.
The accuracy given by the initial 26 PMUs is low (22.7\% correct versus 77.3\% misidentified) due to the very limited number of PMUs and the large number of outage scenarios.
However, as the coverage increases by adding PMUs at high voltage buses, the results have been improved significantly.
When 220 kV+ buses are monitored (868 PMUs), the correctly identified rate without filtering rises up to more than 70\%.
This indicates monitoring high voltage buses is very beneficial and should be considered for future PMU deployments.

\section{Conclusions and Future Work}
\label{sec: conclusions and future work}
The current proposed methods have been implemented and tested on several systems of different scales.
The results show that the proposed identification algorithm using the ac power flow model achieves better identification accuracy compared to the dc approach. 
Additionally, relatively high identification accuracy is achieved with a small number of PMUs.
By using rejection filtering techniques, the misidentified rate can be further reduced, which is crucial for online utilization of the event detection.
The results also demonstrate there are significant benefits of having a higher PMU coverage.
Low number of PMUs makes it difficult to identify the outage on the Ontario system, but future PMU deployments (e.g., with all high voltage buses monitored) exhibit good performance.


%
%


\ifCLASSOPTIONcaptionsoff
  \newpage
\fi



\bibliographystyle{IEEEtran}
\bibliography{loidbib}
\end{document}